\documentclass{emulateapj}
\usepackage{graphicx}
\usepackage[usenames,dvips]{color}
\def\arcsec{$\,^{\prime\prime}$}

\newcommand{\lsim }{{\lower0.8ex\hbox{$\buildrel <\over\sim$}}}
\newcommand{\gsim }{{\lower0.8ex\hbox{$\buildrel >\over\sim$}}}

\def\Chandra{${\it Chandra}$}
\def\HST{${\it HST}$\ }

\def\simge{\mathrel{%
   \rlap{\raise 0.511ex \hbox{$>$}}{\lower 0.511ex \hbox{$\sim$}}}}
\def\simle{\mathrel{
   \rlap{\raise 0.511ex \hbox{$<$}}{\lower 0.511ex \hbox{$\sim$}}}}

\newcommand{\Msun}{\ifmmode {M_{\odot}}\else${M_{\odot}}$\fi}
\newcommand{\Lsun}{\ifmmode {L_{\odot}}\else${L_{\odot}}$\fi}
\newcommand{\Rsun}{\ifmmode {R_{\odot}}\else${R_{\odot}}$\fi}

\shorttitle{A Very Faint X-ray Transient in M15}
\shortauthors{Heinke et al.}

\begin{document}
\title{The Discovery of a Very Faint X-ray Transient in the Globular Cluster M15}  

\author{Craig~O. Heinke\altaffilmark{1,2}, Haldan~N. Cohn\altaffilmark{3}, Phyllis~M. Lugger\altaffilmark{3}}

\altaffiltext{1}{University of Alberta, Dept. of Physics, Room \#238 CEB, 11322 - 89 Avenue, Edmonton, AB, T6G 2G7, Canada; cheinke@phys.ualberta.ca}

\altaffiltext{2}{University of Virginia, Dept. of Astronomy, PO Box 400325, Charlottesville, VA 22903}

\altaffiltext{3}{Astronomy Dept., Indiana University, 727 East 3rd St., Bloomington, IN 47405}

%\slugcomment{}

\begin{abstract}

We have identified an X-ray transient (hereafter M15 X-3) in the globular cluster M15 from an archival \Chandra\ grating observation.  M15 X-3 appears at an X-ray luminosity of $6\times10^{33}$ ergs s$^{-1}$ with a spectrum consistent with an absorbed power law of photon index 1.51$\pm0.14$.
The object is identifiable in archival \Chandra\ HRC-I observations with an X-ray luminosity of $2-6\times10^{31}$ ergs s$^{-1}$ and apparently soft colors, suggesting a neutron star low-mass X-ray binary in quiescence.  We also observe it in outburst in a 2007 \Chandra\ HRC-I observation, and in archival 1994-1995 ROSAT HRI observations.  

We identify a likely optical/UV counterpart with a (possibly transient) UV excess from archival $HST$ data, which suggests a main sequence companion.  
We argue that M15 X-3's behavior is similar to that of the very faint X-ray transients 
which have been observed in the Galactic Center. We discuss several explanations for its very low X-ray luminosity, with the assumption that we have detected its  companion.  
M15 X-3's uniquely low extinction and well-determined distance make it an excellent target for future studies. 

\end{abstract}


%%\maketitle
%----------------------------------------------------------------------

\section{Introduction}\label{s:intro}

Low-luminosity X-ray sources in globular clusters 
were identified with Einstein 
\citep{Hertz83} as a separate class of X-ray sources from 
bright ($10^{36}$$ <$$ L_X$$ <$$ 10^{38}$ ergs/s) LMXBs containing neutron stars (NSs). 
HG83 suggested that these systems were primarily cataclysmic variables (CVs), 
with similar accretion rates to the LMXBs but potential energy wells 
1000 times shallower, but also including some NS LMXBs in quiescence (qLMXBs). 
However, \citet{Verbunt84} argued that the brightest low-$L_X$ systems 
($L_X$$>$$10^{33}$ ergs/s) were too bright to be CVs, and must be qLMXBs.

Many low-$L_X$ cluster sources have now been observed with the 
 {\it Chandra X-ray Observatory}, 
and the answer has been mixed; some of the brighter low-$L_X$ systems 
have been identified with qLMXBs, some with CVs \citep[e.g. in 47 Tuc,][]{Grindlay01a}, while others remain unidentified. 
 The brightest few low-$L_X$ systems 
($5\times10^{33}$$ <$$ L_X$$ <$$ 10^{35}$ ergs/s, generally transient) 
remain a puzzle for either explanation.  These objects are more 
luminous than known CVs, and are brighter than qLMXBs emitting X-rays
from heat stored in the crust between accretion events. 
 However, they are fainter than expected for X-ray 
binaries undergoing outbursts driven by the standard disk instability model 
\citep{King00}.  Eight such systems may have been seen, with various 
X-ray telescopes (Einstein, ROSAT, \Chandra), in several globular clusters, 
but none have been studied in great detail (Heinke et al. 2008, in prep).  
Only one of them \citep[Source B in NGC 6652,][]{Heinke01} shows strong evidence for being a quiescent low-mass X-ray binary, though its accretion history is uncertain.  

Thirty very faint X-ray transients (VFXTs) have been identified, mostly in the Galactic Center region, having peak X-ray luminosities $10^{34}<L_X<10^{36}$ ergs/s and quiescent $L_X$ at least a factor of 10 lower \citep{Muno05,Sakano05,Wijnands06,Degenaar08}.  
Explanation of their very low outburst luminosities and low inferred time-averaged mass transfer rates (\.{M}$\lesssim10^{-12}$ \Msun) may require new paths of binary evolution.
These objects are now being well-studied with \Chandra, XMM-Newton, and $Swift$  through monitoring of the Galactic Center, where the majority have been seen.  
There is some overlap between these VFXTs and the class of low-luminosity X-ray bursters \citep{Cocchi01,Cornelisse02b,Hands04}, which indicates that at least some of these systems are low-mass X-ray binaries.  
Some low-luminosity X-ray bursters show extremely short X-ray outbursts \citep{Wijnands07}, while at least one VFXT, XMMU J174716.1-281048, shows years-long activity at $L_X=$few$\times10^{34}$ ergs/s \citep{delSanto07}.  
We are undertaking a program to search for and study VFXTs in globular clusters, which provide opportunities to study the optical, ultraviolet, and soft X-ray emission not available in observations of the Galactic Center.  

The globular cluster M15 (NGC 7078) is of particular interest for 
X-ray and optical astronomy.  M15 is a massive cluster that shows signs 
of advanced core collapse, with a radial stellar density profile showing 
a steep cusp to within the central arcsecond \citep{Lauer91,Guhathakurta96,SosinKing97}.  
M15 shows a central concentration of relatively underluminous mass, 
that may be composed of 
compact objects \citep[such as heavy white dwarfs and neutron stars,][]{Dull97,Dull03,Baumgardt03} or an intermediate-mass black hole \citep{Gerssen02}.
Its high central density also makes it a likely location for stellar 
encounters, which should produce LMXBs, CVs, and millisecond radio pulsars. 
Studies in the X-ray (requiring the use of \Chandra), optical, and radio have indeed identified two bright LMXBs \citep{White01}, several 
CVs \citep{Charles02,Hannikainen05,Dieball07} and eight radio pulsars \citep{Anderson93}. We use a distance of 10.3 kpc \citep{vandenBosch06} and a column density of  
$N_H=4.6\times10^{20}$ cm$^{-2}$ \citep{Janulis92} in this paper.
Here we report the detection of a very faint X-ray transient in the 
globular cluster M15.  

%\clearpage
\begin{figure}
%\epsscale{0.6}
\includegraphics[angle=0,scale=.47]{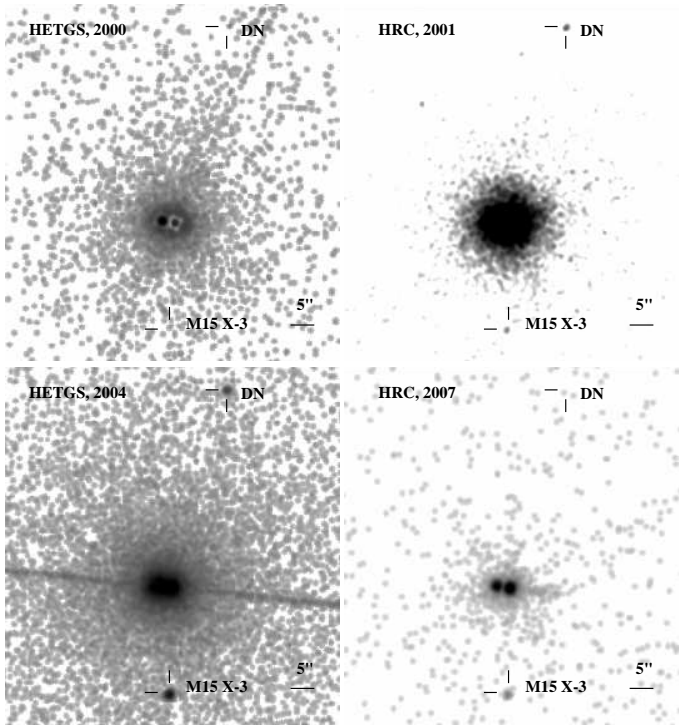}
\caption[M15_ACIS_rev.eps]{ \label{fig:ACIS}
\Chandra\ images of M15, from 2000 ACIS-HETGS, 2001 HRC-I, 2004 ACIS-HETGS, and 2007 HRC-I data.  All data are smoothed with a gaussian of kernel 0.75\arcsec.   The dwarf nova (DN) identified by \citet{Hannikainen05} and M15 X-3 are marked. Readout streaks can be seen at position angles $\sim30$ and $\sim100$ degrees from the saturated LMXBs in the HETGS data.  The greyscale is chosen to emphasize M15 X-3 in each image.
} 
\end{figure}
%\clearpage

\section{Chandra Data Analysis}\label{s:obs}

We reduced all 6 \Chandra\ observations of M15 in the \Chandra\ data archive\footnote{http://cda.harvard.edu/chaser/mainEntry.do} (see Table \ref{tab:obs}).   
All data were reduced and analyzed using CIAO 3.4.1.  
For the ACIS exposures, we generated new bad pixel files, reprocessed the ACIS data removing pixel randomization, and filtered on grade and status.  
For all data, we searched for periods of high, flaring background, but found none.\footnote{A smooth 60\% rise in the background over the length of Obs\_ID 2413 is still small compared to the scattered X-ray background from the LMXBs.}

Several X-ray sources are apparent in the images (Fig. \ref{fig:ACIS}), including the two bright 
LMXBs reported by \citet{White01}, two faint X-ray sources (a dwarf nova, DN, and planetary nebula, PN) reported by 
\citet{Hannikainen05}, and the transient reported here.  
After trying various detection strategies, we chose the following strategies, 
all on 2'$\times$2' images centered on M15's core.
To locate the bright LMXBs in the HETGS observations, we ran the CIAO detection program WAVDETECT  without an energy band restriction.  We did this in order to ensure the bright LMXBs retained as many of their photons as possible, as the zeroth-order images of the two bright LXMBs show substantial pileup.\footnote{Pileup is the recording of more than one photon as part of the same event during a single frametime, which leads to distortion of the energy spectrum and loss of events \citep{Davis01}.}  To detect the other faint X-ray sources in M15 during the ACIS observations, we ran WAVDETECT in the 0.5-7 keV and 0.5-2 keV ranges.    

%\clearpage
\begin{figure}
%\epsscale{0.6}
\includegraphics[angle=0,scale=.47]{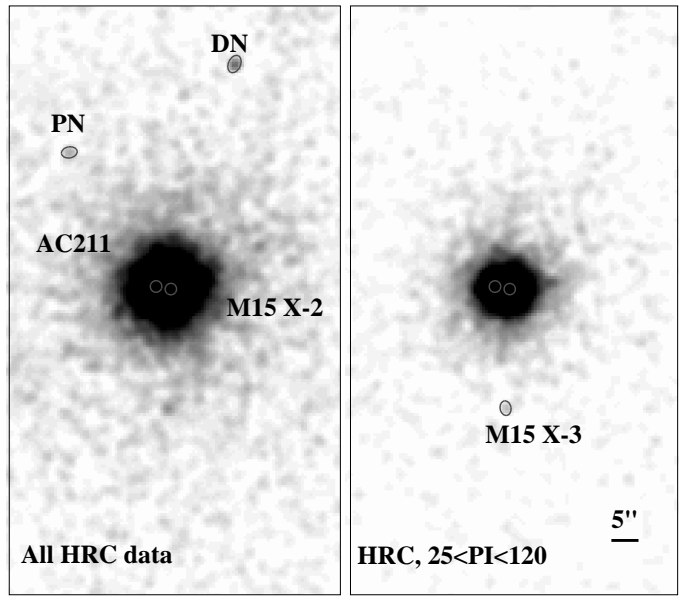}
\caption[M15_HRC_rev.eps]{ \label{fig:HRC}
Combined 2001 \Chandra\ HRC images of M15, including all good events (left) and only a soft 
subset (25$<$PI$<$120, right), binned by 2 and smoothed with a gaussian of kernel 0.75\arcsec.  Ellipses encircle sources detected by WAVDETECT in each image, which are labeled (PN=planetary nebula, DN=dwarf nova).  
} 
\end{figure}
%\clearpage

We reprojected the three 2001 HRC observations and merged them to increase sensitivity. We ran WAVDETECT on two images; the full HRC dataset, and a dataset filtered to include only softer photons by including only events in pulse-invariant (PI) channels between 25 and 120 \citep[cf.][]{Cameron07}.  We binned the HRC data by a factor of two.  
We show both HRC images of M15 in Fig. \ref{fig:HRC}.

We chose WAVDETECT detection thresholds of 1e-6, a conservative limit which should produce much less than 1 false source in each image.  We identified no sources in the 2000 HETGS data besides the two bright LMXBs.  In the combined, full-energy HRC image from 2001 we identified two sources noted by \citet{Hannikainen05}, identified as a dwarf nova (DN, their source C) and a planetary nebula (source D; PN K648), and determined accurate positions for the two LMXBs.
In the 2004 data we identify a new transient (hereafter M15 X-3), as well as the DN.  In the 2007 HRC-I data we again detect the new transient, along with the two bright LMXBs \citep[as reported by][]{Heinke07b}.

Our WAVDETECT run on the filtered HRC data did not detect the PN or DN, but did detect a faint (8-count) source consistent with the position of M15 X-3 (Fig. \ref{fig:HRC}).  No other, spurious, sources were detected in these detection runs, so we believe that we have indeed detected M15 X-3 in quiescence.  Its detection in the filtered HRC data suggests that its quiescent spectrum is very soft, as is typically observed for NS LMXBs in quiescence.  

We aligned our astrometry with the best radio-determined position of AC 211 \citep{Kulkarni90} for the HRC frames, precessing from B1950 to J2000 using the NED precession calculator.\footnote{http://nedwww.ipac.caltech.edu/forms/calculator.html}  We derive our most accurate position for M15 X-3 using the offset from AC 211 in the 2007 HRC-I data, in Table \ref{tab:srclist}.  Positions derived from the marginal HRC-I 2001 detection and 2004 HETGS detection are consistent within 3$\sigma$ with this detection, but are subject to systematic errors (due to the marginal 2001 detection and effects of pileup on the derived position of AC 211 in 2004).  Our listed positions for the other sources are derived from the offsets from AC 211 using the 2001 HRC-I data.

We do not detect sources A and B from \citet{Hannikainen05}, though we do not utilize their psf-subtraction method.   We note that the HRC X-ray source identified with a PN by \citet{Hannikainen05} is undetected in the 2004 HETGS dataset, suggesting that it has a soft spectrum (though not as soft as M15 X-3 in quiescence) or is variable.  We defer further analysis of the other X-ray sources in M15 to future work.  

%\clearpage

\subsection{X-ray Spectral Analysis}\label{s:spec}

Transiently accreting NSs in quiescence are usually seen to have soft,
blackbody-like X-ray spectra, often accompanied by
a harder X-ray component generally fit by a power-law of photon index
1-2 \citep{Campana98a}.  The harder component is of unknown origin;  
an effect of continued accretion, or a shock from a pulsar wind have 
been suggested \citep{Campana98a}.   
 The blackbody-like component is generally understood as the 
radiation of heat from the NS surface. This heat is produced by deep crustal 
heating during accretion, and is radiated by the crust 
on a timescale of $10^4$ years, producing a steady
quiescent thermal NS luminosity \citep{Brown98, Campana98a, Haensel90}. 
At higher luminosities (above $\sim2\times10^{33}$ ergs/s), the spectra of
 faint accreting NSs are generally dominated by the hard component 
\citep[e.g.][]{Jonker04}.  
The spectra of cataclysmic variables can often be represented by one or more 
hot plasma components, e.g. the MEKAL models in XSPEC \citep{Liedahl95}, or, with low statistics, a power-law of photon index 1-2.

%\clearpage
\begin{figure}
%\epsscale{0.6}
\includegraphics[angle=270,scale=.37]{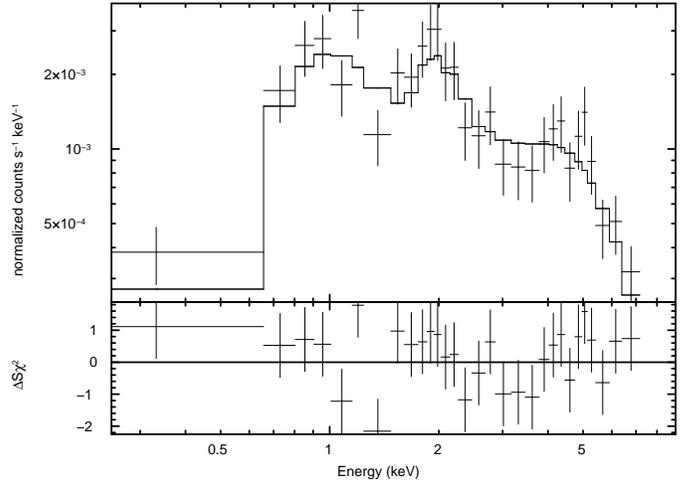}
\caption[M15_X3_spec.eps]{ \label{fig:spectrum}
Top: \Chandra\ 2004 HETGS X-ray spectrum (data and best-fit absorbed power-law model) of M15 X-3.  
Bottom: Residuals to best fit.
} 
\end{figure}
%\clearpage

{\it ACIS:}  
We extract a spectrum, background and response files for M15 X-3 from the 2004 
ACIS data, using the CIAO {\it specextract} script.  
The spectrum can be well fit by an absorbed power-law of photon index 
$\Gamma=1.51\pm0.14$ at the cluster's $N_H$ value (see Table \ref{tab:spec}; Figure \ref{fig:spectrum}).  
Adding a hydrogen-atmosphere neutron star atmosphere model
 \citep[the NSATMOS model,][]{Heinke06a}
does not improve the fit, but we can place constraints on the luminosity 
in a NSATMOS component of $L_X$(0.5-10 keV)$<8.9\times10^{32}$ ergs/s.
A hot thermal plasma model (MEKAL) can also describe the spectrum adequately, 
with a temperature of 14$^{+17}_{-6}$ keV.  

%\clearpage
\begin{figure}
\includegraphics[angle=0,scale=.42]{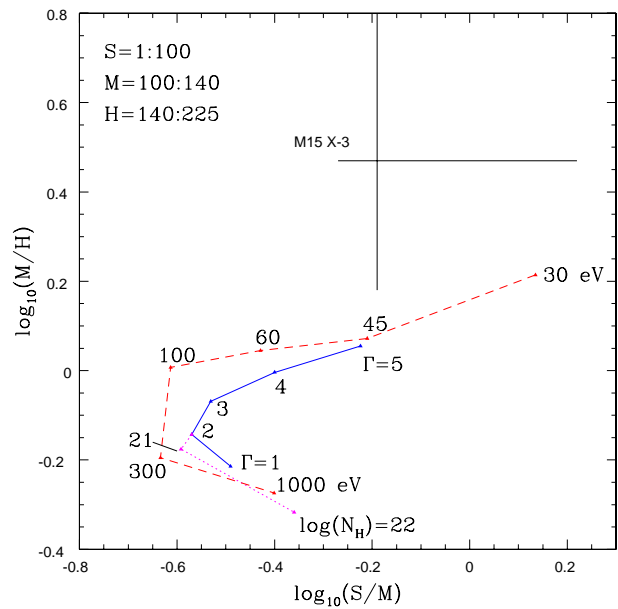}
\caption[M15_X3_spec.eps]{ \label{fig:hrc_color}
HRC-I hardness ratios for various power-law spectra (photon index $\Gamma=1$ to 5, blue solid line) and blackbody spectra (kT=30 to 1000 eV, red dashed line) using the HRC-I response matrix.  We show the effect of increasing the $N_H$ columns from $4.6\times10^{20}$ to $10^{21}$ or $10^{22}$ cm$^{-2}$ for the power-law with photon index 2 (magenta dotted line).  M15 X-3's quiescent colors and 1-$\sigma$ uncertainties are also plotted, suggesting a very soft spectrum. 
} 
\end{figure}
%\clearpage

{\it HRC:}
HRC data is not amenable to standard spectral fitting, but hardness ratios 
can be computed and some spectral information extracted from comparison with the response matrix provided by the Chandra X-ray Center.\footnote{See http://cxc.harvard.edu/cal/Hrc/RMF/}  Only 13 counts are present within a 1\arcsec\ source region in the combined 2001 HRC data, of which $\sim$3 are probably background. Given these limitations, the ratios of counts in PI channels 1:100, 100:140, and 140:255 suggest a power-law photon index $>4$ (Fig. \ref{fig:hrc_color}).  This suggests a very soft source, such as is often seen from quiescent LMXBs \citep{Jonker04}, although we caution that the statistics are very poor and the response matrix uncertain.

We estimate the 0.5-10 keV X-ray luminosity of M15 X-3 in quiescence as $6\times10^{31}$ ergs/s (for a $\Gamma=2$ power-law) to $2\times10^{31}$ ergs/s (for a blackbody of $kT$=110 eV).  A hydrogen atmosphere model \citep[NSATMOS, ][]{Heinke06a} with T=57 eV and 10 km radius provides a soft spectrum and comparable observed countrate, and a bolometric NS luminosity of $8\times10^{31}$ ergs/s.
Such soft X-ray sources would 
be hard to detect in zeroth-order HETGS images; in the 2000 HETGS observation, 
only 2 or 1 counts are expected, given the powerlaw or blackbody spectral models above.  
Therefore the nondetection of M15 X-3 in the 2000 data is not surprising.

\subsection{X-ray Variability}

\begin{figure}
%\epsscale{0.6}
\includegraphics[angle=0,scale=.55]{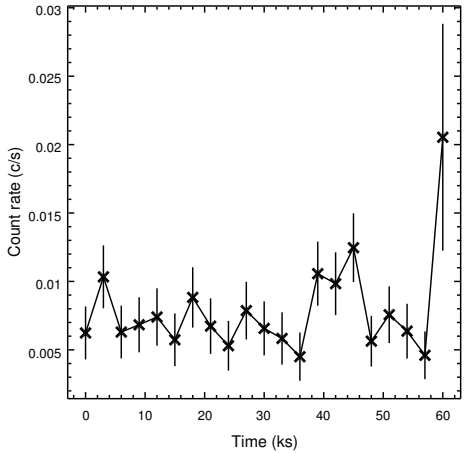}
\caption[lcurve.eps]{ \label{fig:lcurve}
Background-subtracted lightcurve of M15 X-3 during the 2004 ACIS-S/HETGS observation.  
} 
\end{figure}

To assess variability within the 2004 observation, 
we produced barycentered lightcurves of M15 X-3 and analyzed them
 using HEASARC's XRONOS 
software\footnote{http://heasarc.gsfc.nasa.gov/docs/xanadu/xronos/xronos.html}.
We show a lightcurve (binned at 3000 seconds) of M15 X-3 in Fig.\ref{fig:lcurve}.  
A $\chi^2$ test on this lightcurve gives a probability of 0.02 that M15 X-3 is 
constant during the 2004 observation.  The power spectrum shows red 
noise below $3\times10^{-3}$ Hz, but no obvious periodic signals.

%\clearpage
%\clearpage

\section{Archival X-ray Data Analysis}
Several previous X-ray missions have had sufficient resolution to separate M15 X-3 from the other 2 LMXBs in M15 (though not to separate the two bright LMXBs).  We analyze archival data from ROSAT HRI \& PSPC, ASCA SIS, EXOSAT HRI, and Einstein HRI observations to look for evidence of M15 X-3.  First we consider the constraints possible over the past 12 years from RXTE.

\subsection{RXTE All-Sky Monitor Analysis}\label{s:rxte}

We used data from the RXTE All-Sky Monitor\footnote{http://xte.mit.edu} to constrain M15 X-3's flux history.  The known M15 LMXBs (AC 211 and M15 X-2) together contribute roughly $4\times10^{36}$ ergs/s (2-10 keV), and their combined luminosity is seen to be fairly stable to within a factor of two, apart from short periods when M15 is near the Sun and the data quality is poorer \citep{White01}.  Thus, an outburst from M15 X-3 would have to be of similar luminosity to be potentially discernible.  

M15 X-3's observed $L_X$ in 2004 might be the peak of an outburst, or 
 the tail of an outburst with a higher peak $L_X$.  
M15 is not conclusively detected by the ASM during April 2004, at a time when M15 X-3 was active.  A 2-sigma limit to the flux of M15 during the two weeks preceding this observation is $3\times10^{36}$ ergs/s in the 2-10 keV band, assuming a Crab-like spectrum; this is consistent with the X-ray luminosity from M15 over the RXTE era.  A rough limit on the flux earlier in 2004, and over the RXTE era generally, is $6\times10^{36}$ ergs/s.   This suggests that any outbursts from M15 X-3 in the RXTE era did not significantly exceed $3\times10^{36}$ ergs/s, at least for more than a few weeks (but see below).  

Our 2007 HRC-I observation was taken 2 months after the highest fluxes ever 
recorded from M15 by the RXTE ASM (50 mCrab,\footnote{Noted on http://xte.mit.edu/xte\_anno.html} or $1.5\times10^{37}$ ergs/s).  It is possible that these high fluxes were produced by another X-ray source besides the two previously known LMXBs, AC 211 and M15 X-2.  Our HRC-I observation finds M15 X-3 active, but not brighter than in the three other active episodes.  Thus we do not clearly resolve whether M15 X-3 or another source may have produced the high flux episode from M15. 

\subsection{ROSAT Data Analysis}
\begin{figure}
%\epsscale{0.6}
\includegraphics[angle=0,scale=.47]{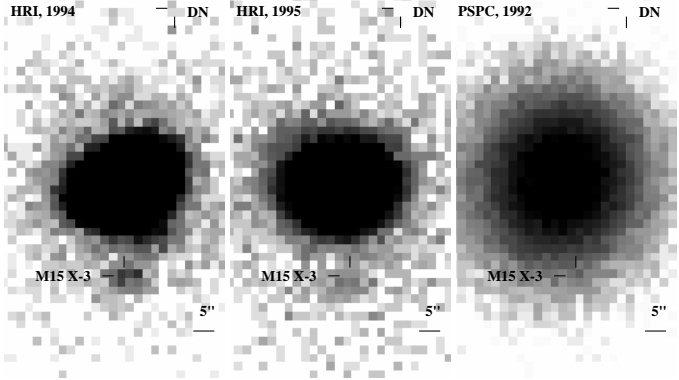}
\caption[M15_ROSATnewer.ps]{ \label{fig:ROS}
ROSAT images of M15.  A weak source can be seen at a position consistent with M15 X-3, 22.6'' south of the two bright LMXBs, in the two HRI images. The image scaling is adjusted for maximum visibility of M15 X-3.
} 
\end{figure}

Two lengthy ROSAT HRI observations of M15 exist in the HEASARC archive (PI: Grindlay), in addition to two shorter PSPC observations (we use only the on-axis one); see Table \ref{tab:obs}.  
The ROSAT XRT/HRI combination has a FWHM of 3'', sufficient to resolve M15 X-3 from the two bright X-ray binaries in the core of M15, and an absolute positional accuracy of 5.3'' \citep{Deutsch98a}. 
 Although our detection algorithms do not identify a source at this location, 
a weak source is visible to the reader in the two HRI images (see Fig. \ref{fig:ROS}, left), with 165$\pm21$ and 35$\pm9$ counts above the local background within 5'' circles.  
Assuming an absorbed power-law spectrum of photon index 1.5 such as seen in 2004, we estimate 
$L_X$(1994, 0.5-10 keV)$\sim7.4\times10^{33}$ ergs/s and 
$L_X$(1995, 0.5-10 keV)$\sim7.7\times10^{33}$ ergs/s.  

%\clearpage
%\clearpage

\subsection{Other HEASARC Data Analysis}
\begin{figure}
%\epsscale{0.6}
\includegraphics[angle=0,scale=.47]{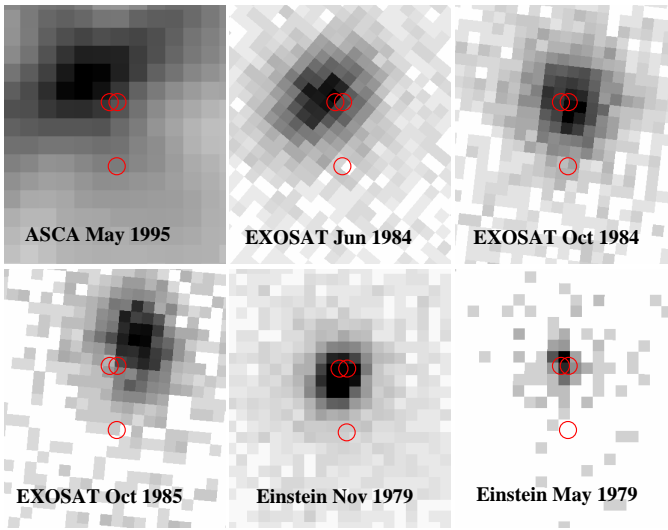}
\caption[M15_HEASARC_c.ps]{ \label{fig:HEA}
Postage stamp images of M15 from six imaging observations (ASCA, EXOSAT, and Einstein).  The positions of both LMXBs and M15 X-3 are indicated with circles of radius 3''.  No boresight corrections have been applied to these images. 
} 
\end{figure}

For the remaining high-resolution imaging X-ray datasets (ASCA SIS, ROSAT PSPC, EXOSAT CMA, and Einstein HRI), we can provide only upper limits to M15 X-3's flux.  The first step is to clarify whether the known LMXBs are those producing the detected source in prior observations.  Figure \ref{fig:HEA} shows images from ASCA, EXOSAT, and Einstein images.  In each case the blind pointing accuracy of the telescope is thought to be much better than 22\arcsec at 90\% confidence,
 and in each case (see below) the detected source is consistent with the positions of the known bright LMXBs, and not with M15 X-3.  

The background is in each case dominated by the point spread functions of the nearby bright LMXBs.  For simple, conservative estimates of the upper limits in each case, we use a simple procedure.  We compare the count rates in 5'' circles around the peak of the observed source (taken to be AC211 and M15 X-2) and the expected location of M15 X-3 (after boresighting to align the detected source with the point midway between the bright LMXBs).  We do not subtract the expected counts from the wings of the LMXBs' point spread function, as the shape of those PSFs are not necessarily well known; this makes our limits very conservative.  We  compute the 0.5-10 keV $L_X$ for the M15 LMXB assuming a powerlaw of photon index 1.51, and multiply by the ratio of the counts in the two circles.  The limits derived are given in Table \ref{tab:obs}.

%\clearpage
%\clearpage

{\bf ASCA}:
  One ASCA observation is available.  The position of a bright ASCA SIS source can be centroided within 12'' after appropriate corrections to the astrometry \citep{Gotthelf00a}.  We find that the position of the M15 X-ray source, according to ASCA, is 21:29:58.88, +12:10:12.97, 12'' NE of the positions of the known LMXBs, and thus consistent.  
%We estimate a conservative upper limit on M15 X-3's flux of 1/10 that of the combined flux of the other LMXBs, or $3\times10^{-11}$ ergs/cm2/s.  

{\bf EXOSAT CMA}:
EXOSAT CMA images from three epochs are available, each using the thin filter.  The EXOSAT CMA provides positions accurate to within 8'' \footnote{http://heasarc.gsfc.nasa.gov/docs/exosat/exosat\_about.html}.  The position of the single detected source is within 8'' of AC211 and M15 X-2 in 2 of the observations, and 12'' to the NW in the third observation. 

{\bf Einstein HRI}:
Einstein HRI images for two epochs are available.  Einstein positions are thought to be accurate to 3.2\arcsec\ \citep{Grindlay84}, and the detected M15 source is consistent with the positions of either bright M15 LMXB. 

%\clearpage
\begin{figure}
\includegraphics[angle=0,scale=.45]{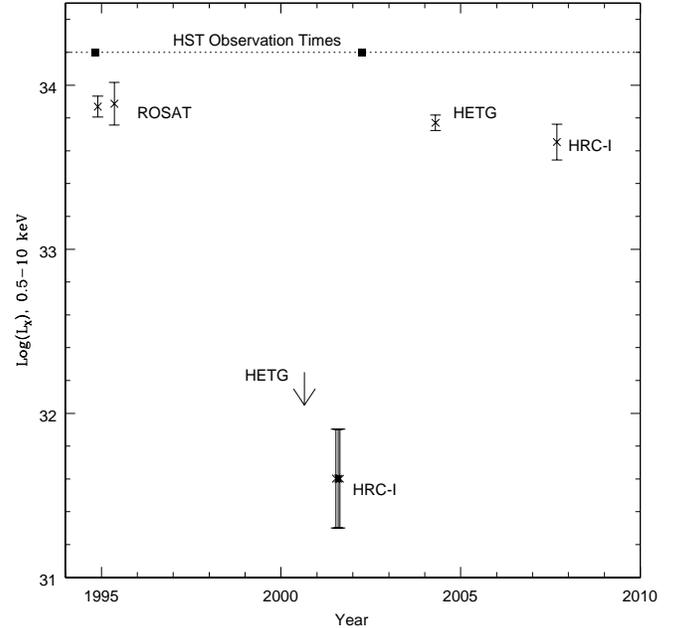}
\caption[timeline.eps]{ \label{fig:timeline}
Timeline of sensitive X-ray detections and upper limits for M15 X-3, from the ROSAT HRI and \Chandra\ HETG and HRC-I imaging.  The times of two useful HST observing epochs are also marked with squares. 
} 
\end{figure}
%\clearpage

\subsection{X-ray Conclusions}\label{sec:xcon}

The ROSAT and \Chandra\ data require at least two substantial changes in flux; a decrease between 1995 and 2000, and an increase between 2001 and 2004 (Fig.\ \ref{fig:timeline}).  Consecutive X-ray observations tend to find the source in the same state, suggesting that typical on and off periods for M15 X-3 are a few years in duration.  Further high-resolution monitoring observations would be greatly useful in constraining M15 X-3's recurrence history.  

Our conclusion from the analysis of archival X-ray datasets on M15 is that if 
M15 X-3 has reached high $L_X$s of $>10^{35}$ ergs/s, it has done so rarely. 
The twelve most sensitive X-ray observations over 29 years all constrain M15 X-3 to $L_X<10^{35}$ ergs/s.  
We consider three scenarios for M15 X-3's mass transfer rate (we assume a 10-km, 1.4 \Msun\ NS accretor for each): a) the observations in hand are representative, and it spends $\sim$half of its time ``on'' at $L_X=6\times10^{33}$ ergs/s; for this case, $\dot{M}\sim2\times10^{-13}$ \Msun/year. 
b) M15 X-3 has undergone outbursts in the past 12 years, which we have happened to miss.  The RXTE ASM coverage indicates no outbursts brighter than $\sim3\times10^{36}$ ergs/s in the past 12 years (apart from the 2007 brightening, \S \ref{s:rxte}), so we assume a maximum $L_X=3\times10^{36}$ ergs/s.  The twelve most sensitive existing X-ray datasets constrain M15 X-3 to $L_X<10^{35}$ ergs/s; so we assume that X-3 is in outburst at $L_X < 3\times10^{36}$ ergs/s only 1/12 of the time.   This gives a mass transfer rate of $\dot{M}<2\times10^{-11}$ \Msun/year.   
c) M15 X-3 has undergone major outbursts, but they were more than 12 years ago; in this case, we cannot constrain its mass transfer rate.

\section{Other Wavelengths}

\subsection{Optical photometry}\label{sec:hst}

\begin{figure}
%\epsscale{0.6}
\includegraphics[angle=0,scale=.45]{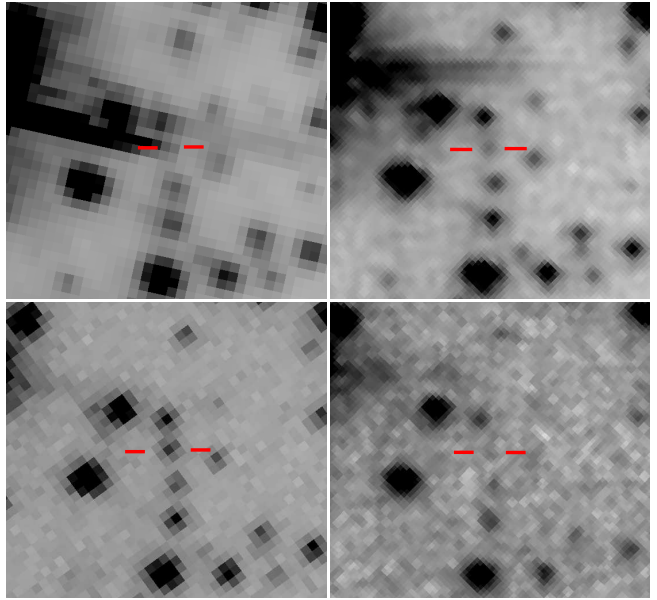}
\caption[M15_transient.ps]{ \label{fig:fchart}
Finding charts for candidate counterpart of M15 X-3.  The location of M15 X-3 is bracketed by (red) dashes in each frame.  Upper left, ACS/WFC F606W image from 2006 (affected by column bleeding); upper right, WFPC2 PC F555W image (2002); lower left, WFPC2 PC F336W image (1994); lower right, WFPC2 PC F439W image (2002).
} 
\end{figure}

 M15 has been the target of numerous \HST\ optical campaigns.  
 M15 X-3's location 20\arcsec\ from the center of M15 reduces crowding, but unfortunately places it outside the field of view of most HST high-resolution instruments during pointings at the core. (The $0.1''$ pixel size of the WFPC2 Wide Field chips  does not adequately resolve
this crowded area.)  Three epochs of imaging cover M15 X-3 with the Planetary Camera (PC) chip, or with ACS/WFC.  Unfortunately, during the ACS observation epoch (2006), the orientation of the spacecraft placed bleeding columns from a nearby bright giant onto M15 X-3's position.  Thus we remain with only two usable epochs, in only three bands: F336W (1994), and F555W and F439W (2002); see Fig. \ref{fig:fchart}, and Table \ref{tab:hst}. 

\begin{figure}
\includegraphics[angle=0,scale=.45]{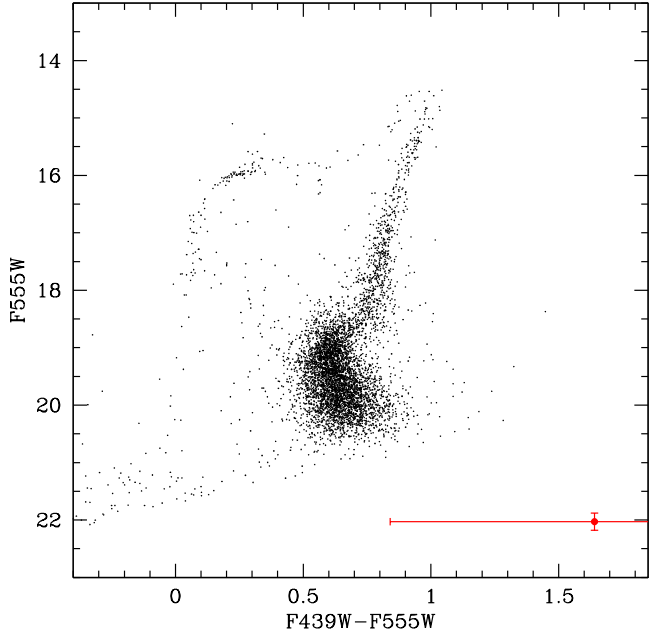}
\caption[m15_x3_bv_cmd.ps]{ \label{fig:bvcmd}
Color-magnitude diagram for M15, in F439W (``$B$'') and F555W
(``$V$'') filters, both from the same observation epoch in 2002.  (The
calibration is described in the text and includes a zeropoint shift of
0.6 mag in F439W to bring it closer to $B$.)  The proposed counterpart
to M15 X-3 is indicated by the red dot with the large error bars to
the red of the main sequence (though consistent with it).  }
\end{figure}

\begin{figure}
\includegraphics[angle=0,scale=.45]{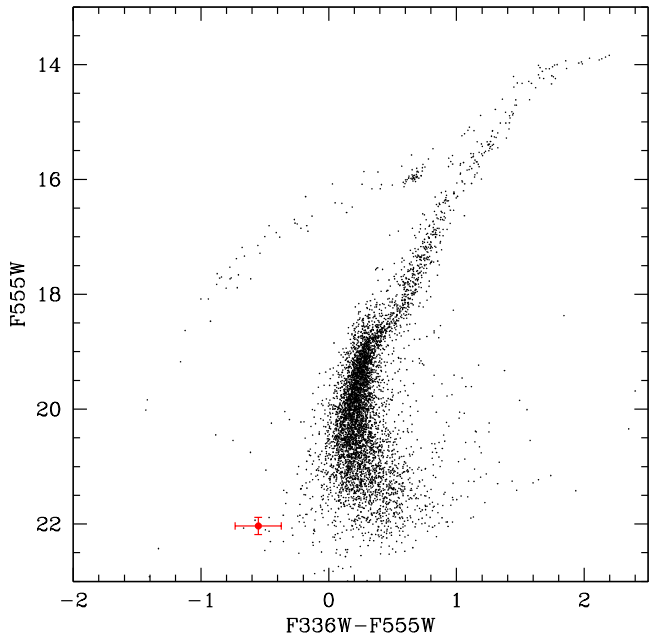}
\caption[m15_x3_uv_cmd.ps]{ \label{fig:uvcmd}
Color-magnitude diagram for M15, in F336W (``$U$'') and F555W
(``$V$'') filters.  (The calibration is described in the text.)  The
proposed counterpart to M15 X-3 is indicated by the red dot with error
bars to the blue of the main sequence.  Note that the F555W and F336W
frames were taken in 2002 and 1994 respectively.  } 
\end{figure}

Since the 2002 data were obtained with
dithering, we constructed drizzled reconstructions of the F555W and
F439W PC fields.  In order to generate approximate color-magnitude
diagrams (CMDs), we performed aperture photometry on the drizzled
frames and on a combined frame of F336W exposures.  We applied the
standard calibration for the STMAG system to the resulting magnitudes.
We then applied an aperture correction to an aperture of $0.5''$.  We
approximately aligned the resulting color-magnitude diagrams (CMDs) to
the Johnson system by adding zeropoint shifts appropriate to a star at
the main-sequence turnoff (MSTO); these shifts are neglible except for
F439W, which was increased by 0.60 mag.  No color-dependent
transformations were employed.  The resulting CMDs are shown in
Figs.~\ref{fig:bvcmd} and \ref{fig:uvcmd}.  Given the use of aperture
photometry without neighbor subtraction, we have not plotted the
photometry for the region of radius $4''$ about the cluster center in
order to produce cleaner fiducial sequences.  Comparison with the
($V$, $B-V$) CMD of \citet{vanderMarel02} shows reasonable alignment
of the magnitude scales in the vicinity of the horizontal branch and
main sequence turnoff.

We use the relative positions of M15 X-3 vs.\ AC211 given in Table 3 to derive a $0.14''$ (2$\sigma$) error circle for M15 X-3's optical counterpart in the HST images. 
We have found a likely counterpart, within $0.05''$ ($<1\sigma$) of the expected position, that shows evidence of possibly transient UV excess associated with phases of higher mass transfer.  
The 2002 data show a clear detection in $V$, but only a marginal detection in $B$ (Fig.~\ref{fig:fchart}), making the $B$ photometry of this object
challenging. Since it was well below the automatic detection threshold
in $B$, it had to be handled as a special case.  The resulting $B-V$
color index puts the object well
 \emph{redward} of the main sequence but with very large errors that include the main sequence (Fig.~\ref{fig:bvcmd}).  As the 2002
      HST data were taken only 8 months after the 2001 HRC
      observations that found M15 X-3 in quiescence, it seems likely
      that it was also in quiescence in 2002.  In contrast,
      comparison of the 1994 $U$ and 2002 $V$ data implies a clear UV excess in
      1994 (see Fig.~\ref{fig:uvcmd}), relative to the main sequence, consistent with a mass transfer event.   
This makes sense, considering that the 1994 HST image was taken only one month before a ROSAT observation that found M15 X-3 to be active (although M15 X-3 could have varied during that month).

%\clearpage
%\vspace*{0.3cm}

\clearpage

%\clearpage

\subsection{Radio images}

The position of M15 X-3 lies in the field of view of four VLA observations reported in the literature.  A 1.4 GHz image by \citet{Kulkarni90} shows no emission from M15 X-3's location, for an upper limit of 0.3 mJy (4$\sigma$) on October 11-12 1988.   A deeper observation at 1.4 GHz between April 9-14, 1990, by \citet{Johnston91} allows an upper limit of $\sim$130 $\mu$Jy (3$\sigma$).  An 8.4 GHz  map from Feb. 11, 1991, by \citet{Knapp96}, finds a marginal 3$\sigma$ detection ($\sim$70 $\mu$Jy) at a position consistent with M15 X-3 (within 2''; the VLA was in hybrid C/D configuration, with a larger beam). A fourth VLA observation on Oct. 13, 2004 at 8.6 GHz, reported by \citet{Bash07}, allows an upper limit of 25 $\mu$Jy (4$\sigma$).

\section{Discussion}

\subsection{Nature of the accretor} 

M15 X-3's position near the center of M15 strongly indicates that it is a cluster member.  M15 X-3's X-ray luminosity measurements fall within a range that is reached by both 
quiescent LMXBs and cataclysmic variables.  
We provide three arguments for why M15 X-3 is more likely to be an accreting neutron star:

{\it Strong X-ray variability}: M15 X-3's X-ray flux declined by two orders of magnitude between 1995 and 2000, and increased again to its prior level between 2001 and 2004.  Such strong X-ray variability is a hallmark of LMXBs, but is rare among most cataclysmic variables.  \citet{Baskill05} observe that nonmagnetic CVs vary by one order of magnitude or less between observations, and do not reach maximum X-ray luminosities above $3\times10^{32}$ ergs/s.  The X-ray luminosities of polars do vary by two orders of magnitude on appropriate timescales, but  they do not reach above $10^{33}$ ergs/s \citep{Ramsay04b}.  Intermediate polars do attain $L_X>3\times10^{33}$ ergs/s, but they rarely show substantial long-term variability \citep{Muno04b}.   

{\it X-ray to optical flux ratio}: The lack of simultaneity complicates this comparison.  However, the 1994 ROSAT HRI observation which found M15 X-3 in its high state was only one month away from the HST WFPC2 F336W observations discussed in \S \ref{sec:hst}.  Considering that M15 X-3 seems to remain in each state for at least a year, we assume that these values can be compared.  M15 X-3's $V$ magnitude in 1994 must have been $>21$ (otherwise it would not be blue during outburst, given the $U$ constraint).  Using log$(F_{Opt})=-0.4V-3.96$, we find $F_X$(0.5-2.5)/$F_{Opt}$=0.5, which is higher than for most CVs \citep{Verbunt97}.  M15 X-3's relatively high X-ray to optical flux ratio suggests a neutron star accretor.   

{\it Variation in X-rays vs. optical}: 
To within a large uncertainty in the $B$ magnitude, M15 X-3 is
    consistent with being on or near the main sequence in the
    ($V$, $B-V$) CMD, and thus is unlikely to be much fainter in
    quiescence.
Our outburst $U$ detection thus limits the total $V$ increase in outburst to one magnitude, while the X-ray flux varies by two orders of magnitude.  Most CVs show larger variations in optical than X-ray \citep[e.g. ][]{Garnavich88}.  

These arguments suggest that M15 X-3 contains an accreting neutron star, although confirmation (through, e.g., X-ray spectroscopy in quiescence) is desirable.  If M15 X-3 contains a neutron star, the observed outbursts are among the faintest seen from neutron star accretors, identifying M15 X-3's behavior as that of a very faint X-ray transient.

%discuss vfxts; 
%estimate recurrence times for transient outburst, 
%estimate mass transfer rate
%discuss King \& Wijnands

\subsection{Very faint X-ray transients}

Very faint X-ray transients (VFXTs) are identified as X-ray transients with peak X-ray luminosities of $L_X=10^{33.5-34}$ to $10^{36}$ ergs/s and 
quiescent luminosities at least 10 times fainter 
\citep{Muno05,Sakano05,Wijnands06}.  
Thirty systems showing such behavior have been identified, with 3-4 of them being active at any moment in the Galactic Center.  
Some rise up to $L_X\sim10^{35}$ ergs/s, with a low ($\lesssim10$\%) duty cycle, while others, such as XMMU J174716.1-281048 and CXOGC J174535.5-290124, maintain $L_X\sim10^{34}$ ergs/s with relatively high ($\sim50$\%) duty cycles \citep{Muno05,delSanto07,Degenaar08}.  
M15 X-3 appears to show similar behavior to the latter group,  being observed to be ``on'' at $L_X\sim6\times10^{33}$ ergs/s in half of the X-ray observations with sufficient sensitivity to detect such a state.  

  Typical, better-understood X-ray transients have peak X-ray luminosities of 
$L_X=10^{36}$ to $10^{38}$ ergs/s, and time-averaged  mass transfer rates of 
\.{M}$\gsim 10^{-11}$ \Msun/year.  The VFXTs, by contrast, show combinations of duty 
cycles and peak $L_X$s which indicate time-averaged mass transfer rates one to two 
orders of magnitude lower than $10^{-11}$ \Msun/year, difficult to understand in a binary evolution context \citep{King06}.  
We note that VFXTs 
may undergo normal outbursts at other (so far unobserved) times, 
and thus might not have particularly low time-averaged mass transfer rates. 
In either case, their unusually low observed accretion luminosities are difficult to 
understand within the standard disk instability model, and thus warrant study.

Several possible explanations for VFXT behavior can be immediately ruled out for M15 X-3 (though they may be relevant for other VFXTs).  Accretion from the wind of a massive (e.g. B) star or a subgiant  \citep{Pfahl02c} can be ruled out for an optically faint star in a globular cluster.  An intermediate-mass black hole accretor \citep{King06} can also be ruled out, due to M15 X-3's offset from M15's core, and the expected rapid settling of massive objects into the core.  

If our observations are representative, we can infer a time-averaged mass transfer rate of $\dot{M}\sim2\times10^{-13}$ \Msun/year (\S\ref{sec:xcon}).
This is roughly the minimum mass-transfer rate achievable through binary evolution of a main-sequence star with a neutron star companion after a Hubble time \citep{King06}, suggesting an extremely low-mass brown dwarf companion might be a possible solution.
However, our HST photometry suggests (with large uncertainties) that the companion to M15 X-3 is a main-sequence star with a mass of $\sim0.65$ \Msun\ \citep[using an isochrone from ][]{Briley04}. 
Confirmation of a main-sequence companion (through deeper high-resolution HST imaging) would clearly rule out an extremely low-mass substellar companion.  
It would also severely constrain possible models of accretion from the wind of a low-mass main sequence star, as \citet{Willems03} found a maximum time-averaged bolometric accretion luminosity of $10^{33}$ ergs/s from stars in this mass range in the pre-LMXB stage.

One plausible alternative for M15 X-3 is an extremely unfavorable inclination angle, such as  
the X-ray transient identified by \citet{Muno05b}, allowing an isotropic X-ray 
luminosity higher by two orders of magnitude.  
At 10 kpc, an outburst reaching $10^{36}$ ergs/s might be predicted to reach 
radio fluxes of 20 $\mu$Jy from a neutron star LMXB, or 1.0 mJy from a black hole LMXB in the hard state \citep[using the correlation from ][]{Migliari06}. 
The stringent Oct. 2004 radio upper limit of 25 $\mu$Jy, 6 months after an X-ray observation showing activity, suggests that if M15 X-3 is highly inclined, it is more likely a neutron star LMXB.    
The possible 70 $\mu$Jy radio detection in 1991, if real, suggests either a major 1991 outburst (unfortunately not covered with sensitive imaging), or that a large fraction of M15 X-3's accretion power at low accretion rates is going into the formation of jets \citep[as found by][ for black hole systems]{Gallo06}.  
 A high-inclination explanation for the low $L_X$ might predict detectable 
eclipses (the 2004 lightcurve (Fig.\ref{fig:lcurve}) excludes eclipses longer than 1000 s), and that the neutron star should not be directly visible in 
quiescence as a blackbody-like X-ray source (testable with ACIS spectroscopy in quiescence).

A second alternative is the removal of most mass transferred from the companion out of the system, perhaps by the pressure of a fast-spinning NS magnetic field, the ``propeller effect''
 \citep{Illarionov75}.  
Detailed simulations of the interaction of an accretion disk with such a spinning magnetic field find that some material does leak down onto the neutron star poles \citep{Romanova05}.  
A moderately effective propeller effect may explain the particularly low mass transfer rates inferred for VFXTs.  
In this case, the dividing line between VFXTs and ``normal'' LMXBs may be ambiguous, dependent on spin rates, magnetic field strength and/or geometry.   VFXTs may slowly accumulate sufficiently massive disks to produce normal outbursts.  
Some objects do show VFXT-like outbursts as well as brighter outbursts, e.g. GRS 1741.9-2853 and XMM J174457-2850.3 \citep{Muno03_GRS,Sakano05,Degenaar08}, supporting this possibility.

A third alternative is variability in the mass transfer rates from the companion, on timescales greater than tens of years, which is thought to occur in cataclysmic variables \citep[e.g.][]{Hameury89}. 
\citep{King06} dismiss mass transfer cycles as an explanation for VFXTs as a class for two reasons: a lack of systems much brighter than the mean mass transfer rate to balance those which are fainter, and a lack of a mechanism that can deliver such strong variability.  
However, the known bright neutron star LMXBs tend to be brighter than evolutionary calculations predict at their orbital periods \citep{Podsiadlowski02,Pfahl03}.  
One very plausible driver of such variability is an irradiation-driven instability, wherein the X-ray irradiation of the secondary causes slow expansion, leading to mass transfer cycles on $\sim10^8$ year timescales \citep{Podsiadlowski91,Hameury93,Ritter00,Buning04}.  

\section{Conclusions}

We have identified a transient X-ray source in the globular cluster M15, denoted M15 X-3.  
This source has been detected at $L_X=4-8\times10^{33}$ ergs/s in 4 ROSAT and Chandra observations, and $L_X\sim2-6\times10^{31}$ ergs/s in 3 Chandra HRC-I observations, with an apparently softer spectrum.  
Archival HST observations reveal a star with an apparent ultraviolet excess at the X-ray position.  Within broad uncertainty, due to a very weak detection in $B$, the location of the proposed counterpart in the ($V$, $B-V$) CMD is
    consistent with it being a main-sequence star.  The X-ray and optical characteristics of this source suggest an accreting neutron star nature, with the X-ray flux history resembling those of ``very faint X-ray transients'' seen in the Galactic Center region.  A 1991 VLA observation may have detected M15 X-3 at $\sim70$ $\mu$Jy.  If M15 X-3 consists of a neutron star accreting from a main-sequence star, as the evidence suggests, the observed accretion rate onto the NS must be rather lower than the expected mass-transfer rate for a Roche-lobe filling system.  

The location of this object is a mixed blessing.  Being located in M15, 20'' from two luminous and persistent X-ray binaries, hides this object from all X-ray instruments but those with the highest angular resolution (only \Chandra\ for the foreseeable future).  Being in such a dense globular cluster also prevents optical followup with any instrument with inferior angular resolution to HST.  However, M15's intrinsic interest has provided us with numerous archival observations that allowed the discovery of M15 X-3.  
%The age of M15 allows constraints on the nature of the accreting system.  
Most importantly, the low extinction in this direction allows future study of this object in quiescence, which is excluded for all other known very faint X-ray transients both in the optical/UV and soft X-rays.  M15 X-3 may be a key object for understanding the nature of very faint X-ray transients.

\acknowledgements

We thank H. Tananbaum and the CXC for granting the DDT observation ObsID 9584 of M15.  
COH warmly thanks M. Muno, R. Wijnands, J. Miller-Jones and T.~J. Maccarone for discussions, and V. Kashyap for HRC-I RMF calculations.   
COH acknowledges the Lindheimer Postdoctoral Fellowship at Northwestern University, Chandra grant G07-8078X to the University of Virginia, and funding from the University of Alberta Physics Department.   
This research has made use of data obtained through the HEASARC online 
service, provided by the NASA/Goddard Space Flight Center. 
RXTE All-Sky Monitor results are provided by the ASM/RXTE team.   
This work is based in part on observations made with the NASA/ESA Hubble Space Telescope, obtained from the data archive at the Space Telescope Science Institute.  
STScI is operated by the Association of Universities for Research in Astronomy, Inc., under NASA contract NAS 5-26555. 

%\vspace*{-1cm}

\bibliography{src_ref_list}
\bibliographystyle{apj}

%tables

%\clearpage

\begin{deluxetable}{lccccr}
\tabletypesize{\footnotesize}
\tablewidth{4.8truein}
\tablecaption{\textbf{Summary of X-ray Observations of M15 X-3}}
\tablehead{
\colhead{OBS\_ID} & \colhead{Obs. Date} & \colhead{Exposure} &
\colhead{Instrument} & \colhead{$L_X$} 
}
\startdata
4879-4882 & 1979 Nov 19 & 5.07 & Einstein-HRI &  $<$3.0e34  \\
656       & 1978 Nov 22 & 0.26 & Einstein-HRI & $<$1e35  \\
657       & 1979 May 17 & 1.60 & Einstein-HRI & $<$3e34  \\
\hline 
28831 & 1984 Jun 30 & 23.73 & EXOSAT-CMA & $<$3e35 \\
37046 & 1984 Oct 22 & 36.15 & EXOSAT-CMA & $<$9e34 \\
63182 & 1985 Oct 20 & 20.27 & EXOSAT-CMA & $<$1.6e35 \\
\hline 
400081N00 & 1992 May 16  &  8.79 & ROSAT-PSPC & $<$3.4e34 \\
400611N00 & 1994 Nov 27  & 52.47 & ROSAT-HRI & $7.4\pm1\times10^{33}$  \\
400611A01 & 1995 May 11  & 27.62 & ROSAT-HRI & $7.7\pm2$e33  \\
\hline
43042000  & 1995 May 16  & 37.38 & ASCA-SIS &  $<$2.7e36 \\
\hline
675  & 2000 Aug 24 & 20.03 & CXO-HETGS & $<$1.8e32 \\
1903 & 2001 Jul 13 &  9.10 & CXO-HRC-I & $2\times10^{31}$ \\
2412 & 2001 Aug 03 &  8.90 & CXO-HRC-I & $2\times10^{31}$ \\
2413 & 2001 Aug 22 & 10.87 & CXO-HRC-I & $2\times10^{31}$ \\
4572 & 2004 Apr 17 & 60.14 & CXO-HETGS & $5.9\times10^{33}$ \\
9584 & 2007 Sep 05 &  2.14 & CXO-HRC-I & $4.5\times10^{33}$ \\
\enddata
\tablecomments{Times in kiloseconds.  
Unabsorbed $L_X$(0.5-10 keV) observed, or 
upper limits reached (see text), in each observation.  We use PIMMS and a powerlaw of photon index 1.5
 absorbed by $N_H=4.6\times10^{20}$ cm$^{-2}$ to estimate detections or upper limits for observations where 
$L_X$ may be $>10^{33}$ ergs/s, or PIMMS and a blackbody of 0.135 keV with the same $N_H$ for observations where $L_X$ is constrained to $<10^{33}$ ergs/s. 
 \label{tab:obs} }
\end{deluxetable}

\begin{deluxetable}{lccccr}
\tabletypesize{\footnotesize}
\tablewidth{4.8truein}
\tablecaption{\textbf{Summary of \HST\ Observations}}
\tablehead{
\colhead{Prog. ID} & \colhead{Epoch} & \colhead{Instrument} &
\colhead{Exposures} & \colhead{Magnitude} 
}
\startdata
5742 & 1994 Oct 26 & WFPC2  & 3$\times$600 F336W & 21.5$\pm$0.2 \\
9039 & 2002 Apr 5  & WFPC2  & 12$\times$16 F555W & 22.0$\pm$0.2  \\
9039 & 2002 Apr 5  & WFPC2  & 4$\times$40  F439W & 23.7$\pm$0.8  \\
%10775 & 2006 May 2 & ACS-WFC & 1x535 F606W &   \\
%10775 & 2006 May 2 & ACS-WFC & 1x615 F814W &   \\
\enddata
\tablecomments{Magnitudes refer to proposed M15 X-3 counterpart,
with calibration as discussed in text.
\label{tab:hst} }
\end{deluxetable}

\begin{deluxetable}{lcccccr}
\tabletypesize{\footnotesize}
\tablewidth{6.6truein}
\tablecaption{\textbf{\Chandra\ Source List}}
\tablehead{
\colhead{Name} & \colhead{RA} & \colhead{Dec} & \colhead{Cts: 2000} &
\colhead{Cts: 2001} & \colhead{Cts: 2004} & \colhead{Cts: 2007}
}
\startdata
AC211 & 21:29:58.310    & 12:10:02.85    & - & 80769(299)  & - & 4735(70)  \\
M15 X-2 & 21:29:58.124(1) & 12:10:02.37(1) & - & 191255(454) & - & 13960(120) \\
DN, src C & 21:29:57.323(6) & 12:10:43.9(1) & [5] & 21(5) & 79(10) & $<1$ \\
PN, K648 & 21:29:59.40(1) & 12:10:27.6(1) & $<2$ & 9(3) & [5] & $<1$  \\
M15 X-3 & 21:29:58.155(3) & 12:09:40.04(6) & $<2$ & 8(3) & 480(23) & 31(6) \\
\enddata
\tablecomments{ 
Positions and counts from WAVDETECT, in 0.5-7 keV band for ACIS data, full band for HRC data. 
 Nondetections indicated with upper limits, subjective detections in 
brackets.  Errors are in parentheses, on the last quoted digit, and include 
only random (not possible systematic) errors. The exposure times for the observations are: 2000 ACIS-S/HETGS, 20.03 ks; 2001 HRC-I, 28.87 ks; 2004 ACIS-S/HETGS, 60.14 ks; 2007 HRC-I, 2.14 ks.
 \label{tab:srclist} }
\end{deluxetable}

%\clearpage
%\begin{landscape}

\begin{deluxetable}{ccccccc}
\tablewidth{7.0truein}
\tablecaption{\textbf{Spectral Fits to M15 X-3, 2004 CXO}}
\tablehead{
\colhead{Model} &  $N_H\times10^{20}$ & $\Gamma$ & 
 $\chi^2_{\nu}$/dof & $L_{X}$ & $L_{X,NS}$  & $kT$ 
 \\
}
\startdata
Pow     & 4.6$^*$ & 1.49$\pm0.13$          & 0.89/29 & $5.9^{+0.4}_{-0.6}\times10^{33}$ & - & -  \\
Pow+NSATMOS & 4.6$^*$ & 1.42$^{+0.18}_{-0.25}$ & 0.90/28 & $5.8^{+0.5}_{-0.6}\times10^{33}$ & $1.7^{+4.7}_{-1.7}\times10^{32}$  & $82^{+26}_{-82}$  \\
MEKAL  & 4.6$^*$ &   -    &    0.96/29 & $5.9^{+0.7}_{-0.5}\times10^{33}$ & - & 14$^{+17}_{-5}$ \\
\enddata
\tablecomments{Power-law, power-law plus neutron star atmosphere, and mekal
 model fits to spectrum of M15 X-3 in 2004.  
Errors are 90\% confidence for a single parameter.  
X-ray luminosities for 0.5-10 keV, in ergs/s.  $kT$ in eV for NSATMOS model, keV for MEKAL model. \\
$^*$: fixed to the assumed cluster value.  
\label{tab:spec}}
\end{deluxetable}

%\clearpage

\end{document}